# Single-photon emission modeling with statistical estimators for the exponential distribution

Artur Czerwinski[1] · Katarzyna Czerwinska[2] · Xiangji Cai[3] · Asad Ali[4] · Hashir Kuniyil[4] · Atta ur Rahman[5] · Saif Al-Kuwari[4] · Saeed Haddadi[6]

**Abstract**
Single-photon sources are used in numerous quantum technologies, from sensing and imaging to communication, making the accurate modeling of their emissions essential. In this work, we propose a statistical framework for describing single-photon emission processes and implement estimators for the exponential distribution to quantify this phenomenon. Our approach provides a reliable method for estimating the radiative decay time, represented by the inverse rate parameter, which is crucial in quantum optics applications. We explore several statistical estimators, including maximum likelihood estimation, minimum-variance unbiased estimator, and best linear unbiased estimator. To validate our theoretical methods, we test the proposed estimators on experimental data, demonstrating their applicability in real-world settings. We also evaluate the performance of these estimators when dealing with censored data, a frequent limitation in photon emission experiments. The analysis allows us to track the performance of the proposed estimators as the amount of available data decreases, providing insights into their reliability for modeling single-photon emission events under limited resources.

**Keywords**  Single-photon emission · Light-matter interaction · Fluorescence · Radiative decay time · Exponential distribution

## 1 Introduction

Precise modeling and control of single-photon emission processes are crucial to the advancement of several quantum technologies, particularly in quantum communication, quantum imaging, and quantum sensing. Single-photon interactions with matter form the basis of many quantum protocols, enabling tasks such as secure quantum communication through quantum key distribution (QKD), or the realization of quantum repeaters for long-distance quantum networks [1–5]. Furthermore, single-photon sources are essential in advanced imaging techniques, such as quantum illumination and ghost imaging, where the quantum properties of light can improve the resolution

---

Extended author information available on the last page of the article and sensitivity beyond classical limits [6, 7]. To harness the full potential of these applications, accurate modeling of photon emission events is necessary to understand the statistical properties of such emissions [8–12]

The photon emission process can often be described by an exponential distribution, particularly in the context of spontaneous emission from atomic or quantum dot systems. In these cases, the exponential distribution provides a probabilistic framework for describing the time intervals between photon emission events, characterized by the radiative decay rate [13, 14]. Estimators based on exponential distribution, such as the maximum likelihood estimator (MLE) [15–17] and best linear unbiased estimator (BLUE) [18], have been applied in various fields, such as photon loss in fiber-based transmission [19]. In the context of photon emission, these estimators play a key role in accurately determining the decay time, which corresponds to the inverse of the rate parameter that characterizes the exponential distribution. Research on this topic is essential for the development of technologies such as heralded single-photon sources [20, 21], efficient quantum memories [22], and quantum sensing [23].

In this paper, we focus on the application of statistical estimators to the modeling of single-photon emission, particularly in systems with complex photon emission behaviors, such as negatively charged nitrogen vacancy (NV$^-$) centers in diamond and other color centers [24–27]. Our goal is to analyze the performance of commonly used estimators, such as MLE and BLUE, within the framework of quantum optics, where precise timing of photon emission is crucial for experimental success. Ultimately, we validate the performance of statistical estimators using experimental data from the study reported in Ref. [13], offering both theoretical insights and practical solutions for modeling and improving single-photon emission processes in quantum systems. Our findings obtained through statistical estimation are compared with other methods reported in the literature. Additionally, we examine the performance of statistical estimators for censored (truncated) data [28], providing insights into the feasibility of single-photon modeling with limited datasets.

The paper is organized as follows: In Sects. 2 and 3, we define the exponential distribution and discuss its relevance to modeling single-photon emission in quantum optics. Section 4 introduces statistical estimators for the exponential distribution, with Sect. 4.1 covering complete-data estimators and Sect. 4.2 focusing on estimators for Type II censored data. In Sect. 5, we demonstrate how the estimation theory can be applied in quantum optics to describe a single-photon emission, which is a two-parameter symmetric process involving both the non-radiative and radiative decay times. The results of this analysis are presented in Sect. 6, where we apply the estimators to experimental data. Finally, the paper concludes in Sect. 7.

## 2 Exponential distribution

In the quantum optics domain, photon emission events are often modeled using the exponential distribution due to their applicability in describing decay processes. The mathematical formulation of the exponential distribution provides a solid foundation for analyzing such stochastic phenomena, especially in quantum systems where decay and emission events are probabilistically distributed over time.

**Definition 2.1** A random variable $X$ is said to follow an exponential distribution with rate parameter $\lambda > 0$, denoted $X \sim \text{Exp}(\lambda)$, if its probability density function (PDF) is given by [17, 29]:

$$f_\lambda(x) = \begin{cases} \lambda e^{-\lambda x}, & x \geq 0, \\ 0, & x < 0, \end{cases} \quad (1)$$

and its cumulative distribution function (CDF) is:

$$F_\lambda(x) = \begin{cases} 1 - e^{-\lambda x}, & x \geq 0, \\ 0, & x < 0. \end{cases} \quad (2)$$

The rate parameter $\lambda$ represents the average rate of photon emission, while its inverse, $\lambda^{-1} \equiv \beta$, which is often called the scale parameter, gives the average time between atom excitation and photon emission events. This is a crucial characteristic when modeling photon emission from quantum systems like atoms or quantum dots. In this context, $\beta$ is often referred to as the radiative decay time. The expected value and variance of the exponential distribution, given by

$$\mathbb{E}[X] = \frac{1}{\lambda} \quad \text{and} \quad \text{Var}[X] = \frac{1}{\lambda^2}, \quad (3)$$

can be used to predict the timing and spread of photon emissions in such systems.

In many quantum optics experiments, moment-generating functions (MGFs) provide insights into higher-order moments, allowing for the calculation of quantities like $\mathbb{E}[X^k]$. Using the MGF, we can analyze the statistical properties of photon emission events. For the exponential distribution, the MGF is given by:

$$m_X(t) = \mathbb{E}[e^{tX}] = \frac{\lambda}{\lambda - t} \quad \text{for} \quad t < \lambda. \quad (4)$$

The $k$-th raw moment of the distribution can be derived by differentiating the MGF $k$ times with respect to $t$:

$$\mu_k = \frac{\partial^k}{\partial t^k} m_X(t) \bigg|_{t=0} = \frac{k!}{\lambda^k}. \quad (5)$$

These moments quantify the time intervals between successive photon emissions, which are crucial in both theoretical and experimental studies of quantum optical systems. The first moment $k = 1$ represents the mean photon emission time, while higher-order moments $k > 1$ characterize the variability and dispersion of emission times [30].

## 3 Single-photon emission

### 3.1 Photon emission as a memoryless process

An essential feature of the exponential distribution, particularly relevant in quantum optics, is its *memorylessness* [17]. This property implies that the probability of a photon being emitted at a certain time is independent of how much time has already elapsed since the excitation.

**Theorem 3.1** *Formally, for a random variable $X \sim \text{Exp}(\lambda)$, the memorylessness property is expressed as:*

$$P(X > x_1 + x_2 | X > x_1) = P(X > x_2), \quad \forall x_1, x_2 \geq 0. \tag{6}$$

*Proof* The left-hand side of Eq. (6) can be rewritten as:

$$\begin{aligned}
P(X > x_1 + x_2 | X > x_1) &= \frac{P(X > x_1 + x_2 \cap X > x_1)}{P(X > x_1)} \\
&= \frac{P(X > x_1 + x_2)}{P(X > x_1)} \\
&= \frac{e^{-\lambda(x_1+x_2)}}{e^{-\lambda x_1}} \\
&= e^{-\lambda x_2} \equiv P(X > x_2).
\end{aligned}$$

□

The above derivation confirms that the probability of an atom surviving in the excited state beyond a certain time interval is independent of how long it has survived in the past, reinforcing the spontaneous nature of emission events. This memorylessness is also reflected in the quantum mechanical description of single-photon emission, highlighting an inherent symmetry in the process. Specifically, if no photon has been emitted up to a given time $x_1$, the probability of emitting a photon in the future remains governed by a simple exponential distribution, independent of $x_1$. This symmetry ensures that the emission dynamics is Markovian, with the system's future behavior determined solely by its present state. In quantum optics, this property can be conceptually linked to the wave function collapse upon measurement: the detection of a photon defines the system's state, and subsequent evolution proceeds independently of prior temporal evolution. This statistical symmetry between past observations and future predictions forms a foundational principle for analyzing photon emission processes.

### 3.2 Photon emission as a stochastic process

Photon emission in quantum optics can be treated as a stochastic process, where the time between excitation and photon emission follows an exponential distribution. In the quantum regime, this can describe phenomena such as spontaneous emission, where the photon emission probability over time decays exponentially. This is a direct consequence of the underlying physical mechanisms, such as the interaction between an atom and the electromagnetic field. In such models, the probability that a photon is emitted within a certain time interval is directly governed by the rate parameter $\lambda$, which reflects the inherent decay rate of the system.

Furthermore, the quantum state describing the presence or absence of a photon at any given time can be represented within the Fock space formalism [31]. Consider a simple model, where $\rho(t)$ denotes the Fock state at time $t$, leading to a decomposition described by:

$$\rho(t) = e^{-\lambda t}|0\rangle\langle 0| + (1 - e^{-\lambda t})|1\rangle\langle 1|. \tag{7}$$

Here, $|0\rangle$ represents the vacuum state (no photon emission), and $|1\rangle$ represents the single-photon state emitted by the atom. The probability of staying in the excited state (no photon emission) diminishes exponentially over time due to the decay process, while the probability of emitting a photon increases asymptotically to one, as described by the CDF.

## 4 Statistical estimators for the exponential distribution

Statistical estimators play a crucial role in inferential statistics, enabling researchers to make estimates about population parameters based on sample data. Given a set of $n$ observations, denoted by $X_1, X_2, \ldots, X_n$, a statistical estimator is a rule or method that transforms the data into an estimate of the unknown parameter. These estimators, denoted generally as $\hat{\theta}$, are functions that map sample outcomes from the sample space $\Omega$ to the parameter space $\Theta$. They are designed to provide a best guess of the true parameter $\theta$, often subject to properties such as unbiasedness, consistency, and efficiency [16, 32].

In general, estimators can be divided into various categories, depending on the method used to derive them. Some common methods include the method of moments estimator, maximum likelihood estimator, and the minimum-variance unbiased estimator. In this section, we focus on specific estimators suited for the exponential distribution, which can be applied to the study of photon emission events in quantum optics [15, 33].

### 4.1 Complete-data estimators for the exponential distribution

The exponential distribution, characterized by the rate parameter $\lambda$, is widely used to model the time between events in a Poisson process. Several estimators can be applied to estimate the parameter $\lambda$ based on observed data. In the following, we outline the commonly used estimators for this distribution.

In this context, complete-data estimators refer to methods based on a dataset $X_1, X_2, \ldots, X_n$, where all observed events are fully recorded without truncation or censoring. This ensures that statistical estimators can be applied directly without corrections for missing or incomplete observations.

### 4.1.1 Method of moments estimator

The method of moments provides a straightforward approach to parameter estimation by equating sample moments to theoretical moments [32]. For the exponential distribution, the first moment (mean) is known to be $\mathbb{E}[X] = 1/\lambda$. Given a sample of size $n$, with sample mean $\overline{X}$, the method of moments estimator for $\lambda$ is:

$$\hat{\lambda}_{\text{ME}} = \frac{1}{\overline{X}}, \tag{8}$$

where $\overline{X} = (X_1 + \ldots + X_n)/n$ and the lower subscript "ME" stands for the moment estimator. This estimator is simple to compute and provides a reasonable approximation of the true parameter when the sample size is sufficiently large.

### 4.1.2 Maximum likelihood estimator

Maximum likelihood estimation (MLE) is another popular method that maximizes the likelihood function, which measures the plausibility of the parameter given the observed data. For the exponential distribution, the likelihood function based on $n$ independent observations $X_1, X_2, \ldots, X_n$ is:

$$\mathcal{L}(\lambda) = \prod_{i=1}^{n} \lambda e^{-\lambda X_i} = \lambda^n \exp\left(-\lambda \sum_{i=1}^{n} X_i\right). \tag{9}$$

Maximizing this function with respect to $\lambda$ leads to the MLE for $\lambda$ [19], given by:

$$\hat{\lambda}_{\text{MLE}} = \frac{1}{\overline{X}}, \tag{10}$$

which, interestingly, coincides with the method of moments estimator for the exponential distribution. Since the exponential distribution is fully characterized by its mean, both MLE and the method of moments rely on $\overline{X}$, resulting in identical estimators.

Let us now consider the variance of the MLE for the rate parameter $\lambda$. Next, we derive the variance for the scale parameter $\beta = 1/\lambda$.

To compute the variance of $\hat{\lambda}_{\text{MLE}}$, we first note that the sample mean $\overline{X}$ follows a gamma distribution with shape parameter $n$ and rate parameter $\lambda$, i.e., $\overline{X} \sim \text{Gamma}(n, \lambda)$. The expected value and variance of $\overline{X}$ are:

$$\mathbb{E}[\overline{X}] = \frac{1}{\lambda}, \quad \text{Var}(\overline{X}) = \frac{1}{n\lambda^2}. \tag{11}$$

Since $\hat{\lambda}_{\text{MLE}} = 1/\overline{X}$, we use the delta method to approximate the variance of $\hat{\lambda}_{\text{MLE}}$. Applying the delta method, one gets:

$$\text{Var}(\hat{\lambda}) \approx \left[\frac{\partial}{\partial \bar{x}}\left(\frac{1}{\overline{X}}\right)\right]^2 \text{Var}(\overline{X}) = \frac{1}{\overline{X}^4} \cdot \text{Var}(\overline{X}). \quad (12)$$

Substituting $\text{Var}(\overline{X}) = (n\lambda^2)^{-1}$, we obtain:

$$\text{Var}(\hat{\lambda}_{\text{MLE}}) = \frac{1}{n}. \quad (13)$$

Thus, the variance of the MLE for the rate parameter $\lambda$ is inversely proportional to the sample size $n$.

Now, we consider the scale parameter $\beta = 1/\lambda$. The MLE for $\beta$ is the sample mean:

$$\hat{\beta}_{\text{MLE}} = \overline{X}, \quad (14)$$

which is the inverse of Eq. (10).

Based on Eq. (14), we notice that the variance of $\hat{\beta}_{\text{MLE}}$ is simply the variance of the sample mean. As mentioned earlier, the variance of $\overline{X}$ is:

$$\text{Var}(\overline{X}) = \frac{1}{n\lambda^2}. \quad (15)$$

Substituting $\lambda = 1/\beta$ (so that $\lambda^2 = 1/\beta^2$), we get:

$$\text{Var}(\hat{\beta}_{\text{MLE}}) = \frac{\beta^2}{n}. \quad (16)$$

Thus, the variance of the MLE for the scale parameter $\beta$ depends on both the sample size $n$ and the true value of $\beta$.

To assess the statistical efficiency of this estimator, we examine the Fisher information for a single observation. For the exponential distribution parameterized by the scale parameter $\beta$, the Fisher information is:

$$\mathcal{I}_1(\beta) = \mathbb{E}\left[\left(\frac{\partial}{\partial \beta}\log f(X;\beta)\right)^2\right] = \frac{1}{\beta^2}. \quad (17)$$

For $n$ independent observations, the total Fisher information is

$$\mathcal{I}_n(\beta) = \frac{n}{\beta^2}. \quad (18)$$

The Cramér–Rao lower bound (CRLB) for the variance of any unbiased estimator of $\beta$ is then

$$\text{CRLB} = \frac{1}{\mathcal{I}_n(\beta)} = \frac{\beta^2}{n}. \tag{19}$$

Since the variance of the MLE in Eq. (16) exactly attains this bound, the estimator $\hat{\beta}_{\text{MLE}}$ is said to be efficient.

Statistical efficiency is defined as the ratio of the CRLB to the actual variance of the estimator:

$$\text{Eff} = \frac{\text{CRLB}}{\text{Var}(\hat{\beta}_{\text{MLE}})}. \tag{20}$$

Substituting values yields

$$\text{Eff} = \frac{\beta^2/n}{\beta^2/n} = 1. \tag{21}$$

This confirms that the MLE estimator for the scale parameter $\beta$ is fully efficient.

### 4.1.3 Minimum-variance unbiased estimator

The minimum-variance unbiased estimator (MVUE) for the exponential distribution can be derived using the Lehmann–Scheffé theorem [34, 35], which ensures that the estimator achieves the lowest possible variance among all unbiased estimators. For the exponential distribution, by following the Lehmann–Scheffé theorem, we can prove that the estimator

$$\hat{\lambda}_{\text{MVUE}} = \frac{n-1}{\sum_{i=1}^{n} X_i} = \frac{n-1}{n\overline{X}}, \tag{22}$$

is a minimum-variance estimator since it is a function of a sufficient and complete statistic, i.e., $\hat{\lambda}_{\text{MVUE}} \sim \text{MVUE}[\lambda]$ [19].

One can also compute the variance of $\hat{\lambda}_{\text{MVUE}}$ and obtain

$$\text{Var}(\hat{\lambda}_{\text{MVUE}}) = \frac{\lambda^2}{n-2}. \tag{23}$$

To find the variance of the scale parameter, we can utilize the delta method. The variance of $\hat{\beta}$ can be approximated as follows:

$$\text{Var}(\hat{\beta}_{\text{MVUE}}) \approx \left[ \frac{\partial}{\partial \lambda} \left( \frac{1}{\lambda} \right) \right]^2 \text{Var}(\hat{\lambda}_{\text{MVUE}}). \tag{24}$$

The derivative of $\lambda^{-1}$ with respect to $\lambda$ is given by:

$$\frac{\partial}{\partial \lambda}\left(\frac{1}{\lambda}\right) = -\frac{1}{\lambda^2}. \tag{25}$$

Substituting this into the variance expression, we have:

$$\text{Var}(\hat{\beta}_{\text{MVUE}}) = \left(-\frac{1}{\lambda^2}\right)^2 \cdot \text{Var}(\hat{\lambda}_{\text{MVUE}})$$
$$= \frac{1}{\lambda^2} \cdot \frac{\lambda^2}{n-2} = \frac{1}{n-2}.$$

Thus, the variance of the MVUE for the scale parameter $\beta$ is expressed as:

$$\text{Var}(\hat{\beta}_{\text{MVUE}}) = \frac{1}{n-2}. \tag{26}$$

Although the MVUE is designed to minimize the variance for the rate parameter $\lambda$, it is important to consider its performance when estimating the corresponding scale parameter $\beta = 1/\lambda$. The MVUE provides the lowest possible variance for $\lambda$, ensuring that any unbiased estimator of $\lambda$ cannot have a lower variance. However, when this estimator is transformed to obtain $\beta$, the nonlinear relationship between $\lambda$ and $\beta$ leads to a potential increase in the variance of the $\beta$-estimator.

In particular, the variance of the MVUE for $\beta$ can exceed that of the MLE method, which is known to provide efficient estimators that achieve the lowest possible variance (Cramér–Rao lower bound [36]), see Eq. (21). For this reason, while MVUE is theoretically optimal for $\lambda$, it may not offer the same advantage when estimating $\beta$, compare Eq. (16) with Eq. (26). In the context of modeling the decay times in single-photon emission processes, where $\beta$ represents the decay time, it is more practical to rely on the MLE approach, as it provides a more efficient estimate of $\beta$ with a lower variance.

Thus, despite the appeal of MVUE for $\lambda$, we refrain from applying it to $\beta$ in this work. Instead, we prioritize MLE estimators, which are more suitable for achieving the most accurate and reliable decay time estimates in our single-photon emission modeling framework.

### 4.2 Estimators for type II censored data

Censored data occur when the sample of observations is only partially known. Type II censoring specifically refers to a scenario in which a random sample consists of observations up to a certain point, where only the the first $r$ order statistics are observed while the remaining $n-r$ observations are not [28]. This setup often arises in experimental conditions where only a subset of data can be collected due to various limitations, such as time constraints or resource availability.

Let $X_1, \ldots, X_n$ be a random sample of the same distribution. We obtain Type II censored data if we observe exactly $r$ first results (naturally, $r \leq n$), assuming that the observations are arranged in non-decreasing order. In other words, Type II censored data consist of the $r$ consecutive order statistics: $X_{1:n}, \ldots, X_{r:n}$ [37–40].

In the following, we explore different estimators tailored for Type II censored data from an exponential distribution.

### 4.2.1 Maximum likelihood estimator

The MLE is a popular method for estimating parameters by maximizing the likelihood function. For Type II censored data, the likelihood function is derived from the joint density of the observed order statistics. The MLE for the rate parameter $\lambda$ is obtained by differentiating the logarithm of the likelihood function and solving for $\lambda$. The resulting estimator $\hat{\lambda}_{\mathrm{MLE}}(r)$ utilizes the observed values to provide a consistent estimation of the parameters

$$\hat{\lambda}_{\mathrm{MLE}}(r) = \frac{r}{T_r} \quad \text{and} \quad \hat{\beta}_{\mathrm{MLE}}(r) = \frac{T_r}{r}, \tag{27}$$

where $T_r$ is a statistic for the Type II censored data defined as

$$T_r = \sum_{i=1}^{r} X_{i:n} + (n-r) X_{r:n}. \tag{28}$$

Under the assumption that the full sample is used (i.e., $r = n$), the estimator $\hat{\beta}_{\mathrm{MLE}}(r)$ in Eq. (27) simplifies to:

$$\hat{\beta}_{\mathrm{MLE}}(n) = \frac{1}{n} \sum_{i=1}^{n} X_{i:n} = \overline{X}, \tag{29}$$

which is equivalent to the complete-data MLE for the exponential distribution, cf. Equation (14).

The variances corresponding to both parameters can be computed analogously as in Eqs. (13) and (16):

$$\mathrm{Var}[\hat{\lambda}_{\mathrm{MLE}}(r)] = \frac{1}{r} \quad \text{and} \quad \mathrm{Var}[\hat{\beta}_{\mathrm{MLE}}(r)] = \frac{\beta^2}{r}. \tag{30}$$

### 4.2.2 Best linear unbiased estimator

In fluorescence modeling, the radiative decay time of photon emission is an essential parameter, often corresponding to the inverse of the rate parameter $\lambda^{-1} \equiv \tau_R$ of an exponential distribution. One of the approaches to estimate this parameter is the best linear unbiased estimator (BLUE), which, for order statistics, takes the form of a linear combination of the ordered sample values $X_{1:n}, \ldots, X_{r:n}$ [18, 19, 41]

$$\hat{\beta}_{\text{BLUE}}(r) = \sum_{j=1}^{r} c_j X_{j:r}, \tag{31}$$

where the coefficients $c_j$ are relative weights adjusted to ensure the BLUE properties.

The starting decomposition of the estimator $\hat{\beta}_{\text{BLUE}}(r)$ Eq. (31) makes it optimally tailored for the need to determine the decay time. It can be proved that the BLUE for estimating $\beta$ based on the first $r$-order statistics is given by [18, 19]:

$$\hat{\beta}_{\text{BLUE}}(r) = \frac{T_r}{r}, \tag{32}$$

which is equivalent to Eq. (27).

This estimator minimizes the variance among all unbiased linear estimators of $\beta \equiv \lambda^{-1}$, providing an efficient estimation method for the radiative decay time in fluorescence experiments.

Therefore, starting from the assumptions of BLUE, we arrive at the MLE in the case of censored data as well as when the full sample is considered, making both estimators equivalent in this case.

### 4.2.3 Minimum-variance unbiased estimator

The MVUE is a statistic that achieves the lowest variance among all unbiased estimators of the parameter $\lambda$. For Type II censored data, the MVUE for the rate parameter $\lambda$ can be derived using the sufficient statistic that incorporates the observed order statistics. The construction of the MVUE leverages the properties of the order statistics and their distributional characteristics. The estimator $\hat{\lambda}_{\text{MVUE}}(r)$ is based on the formulation of the statistic incorporating the observed values from the order statistics

$$\hat{\lambda}_{\text{MVUE}}(r) = \frac{r-1}{T_r} \quad \text{and} \quad \hat{\beta}_{\text{MVUE}}(r) = [\hat{\lambda}_{\text{MVUE}}(r)]^{-1}. \tag{33}$$

One can also compute the variance of $\hat{\lambda}_{\text{MVUE}}(r)$ and obtain

$$\text{Var}[\hat{\lambda}_{\text{MVUE}}(r)] = \frac{\lambda^2}{r-2}, \tag{34}$$

and

$$\text{Var}[\hat{\beta}_{\text{MVUE}}(r)] = \frac{1}{r-2}. \tag{35}$$

As the number of uncensored observations $r$ decreases, the variance for $\beta$ grows, making the estimation even less reliable than in the case of complete-data estimation.

## 5 Statistical estimators to single-photon emission

In the context of modeling realistic photon emission processes, where both non-radiative and radiative decays are involved, we assume that the total decay time can be modeled as the sequential occurrence of two independent processes. The first process is a fast non-radiative decay with characteristic time $\tau_N = \lambda_N^{-1}$, and the second process is a radiative decay with characteristic time $\tau_R = \lambda_R^{-1}$. These two processes are independent, and the system decays first through the non-radiative channel, followed by the radiative emission. Consequently, the total observed decay time is described by the sum of two exponential distributions, one for each decay process.

While non-radiative decay does not lead to photon emission, radiative decay is associated with the emission of a photon. As the system undergoes these decays in sequence, the PDF for the observed decay time, $p(t)$, can be derived as a convolution of the two exponential distributions, $p_N(t)$ and $p_R(t)$, corresponding to the non-radiative and radiative decay paths, respectively. The result is a hypoexponential distribution, which appropriately captures the sequential nature of these two decay processes [13].

The PDF for the total decay time, $p(t|\lambda_N, \lambda_R)$, is given by the following expression:

$$p(t|\lambda_N, \lambda_R) = \int_0^\infty p_N(t') p_R(t-t') dt' \qquad (36)$$
$$= \frac{\lambda_N \lambda_R}{\lambda_R - \lambda_N} \left( e^{-\lambda_N t} - e^{-\lambda_R t} \right),$$

where $t$ represents the observed decay time, and $\lambda_N \neq \lambda_R$ ensures that the convolution has a well-defined form. This model allows us to describe the overall decay process, including both the non-radiative and radiative components, while accurately capturing the behavior of photon emission in systems where these two processes occur sequentially [13, 14].

In the special case when $\lambda_N = \lambda_R = \tilde{\lambda}$, the above expression becomes undefined due to division by zero. However, the convolution of two exponential distributions with identical rate parameters is well-defined and results in a gamma distribution with shape parameter $k = 2$ and scale parameter $\beta = 1/\tilde{\lambda}$. The corresponding PDF is:

$$p(t, \tilde{\lambda}) = \tilde{\lambda}^2 t e^{-\tilde{\lambda} t}, \qquad (37)$$

which is consistent with the general case and provides a physically meaningful description of the total decay time. This special case representation has been incorporated to ensure that the exponential model for fluorescence remains valid even when the rates of radiative and non-radiative decay are equal.

The hypoexponential distribution provided in Eq. (36) reflects a mathematical symmetry between the non-radiative and radiative decay processes. While these processes differ physically, their contributions to the overall observed decay time are unified in a probabilistic framework, highlighting the complementary roles they play in determining the fluorescence dynamics

Estimating both $\lambda_N$ and $\lambda_R$ requires solving coupled likelihood equations derived from the hypoexponential model [42]. Consequently, standard MLE techniques become impractical for large datasets due to computational complexity. As an alternative, we propose a simplified approach that assumes the non-radiative time $\tau_N$ is small compared to the radiative time $\tau_R$, i.e., $\tau_N \ll \tau_R$ (or equivalently $\lambda_N \gg \lambda_R$). This assumption allows us to focus solely on $\tau_R$, significantly simplifying the estimation process. Under this condition, we can approximate the joint probability as:

$$\lim_{\lambda_N \to +\infty} p(t|\lambda_N, \lambda_R) = \lambda_R e^{-\lambda_R t}. \tag{38}$$

The result in Eq. (38) is equivalent to the standard exponential distribution with a single parameter $\lambda_R = \tau_R^{-1}$. This implies that, when we consider the non-radiative time negligible, the distribution of the overall decay time reduces to an exponential form, dominated by the radiative decay process.

The approximation of neglecting the non-radiative time is a reasonable assumption in systems such as nitrogen-vacancy (NV) centers, where experimental data often suggest that the non-radiative process occurs over a narrow time window, leading to minimal variance in observed non-radiative decay times. More specifically, in crystals non-radiative relaxation typically occurs on a picosecond timescale, whereas radiative decay lasts several nanoseconds [43]

Additionally, the approximation that $\tau_N \ll \tau_R$ is often supported by physical considerations. Non-radiative processes generally involve faster transitions, such as phonon-mediated processes or internal conversions, which complete on much shorter timescales than photon emission. As a result, their contribution to the overall decay process can be effectively "washed out" when combined with the slower radiative decay, justifying the approximation of Eq. (38).

The primary advantage of this approach is that it allows the application of standard estimators for the exponential distribution, bypassing the need for more complex methods typically required for convolutions of two different distributions. This reduction in computational complexity is particularly valuable when dealing with large datasets or when experimental conditions limit the resolution of non-radiative decay times.

## 6 Results for experimental data from NV⁻ centers in a diamond

The experimental data analyzed in this study originate from a heralded single-photon source (HSPS) experiment utilizing NV⁻ centers in diamond, as described in Ref. [13]. The experiment was designed to address the challenge of nonresonant excitation by employing a spectrally broad absorber, such as a color center in diamond, where phononic sidebands enable excitation without the need for narrow bandwidth light [24]. In this setup, a heralded single photon generated by spontaneous parametric down-conversion (SPDC) is absorbed by the NV⁻ center, followed by the emission of another photon. The arrival time of the emitted photon is then measured using time-resolved single-photon detection techniques, allowing precise tracking of the fluorescence decay dynamics of the system.

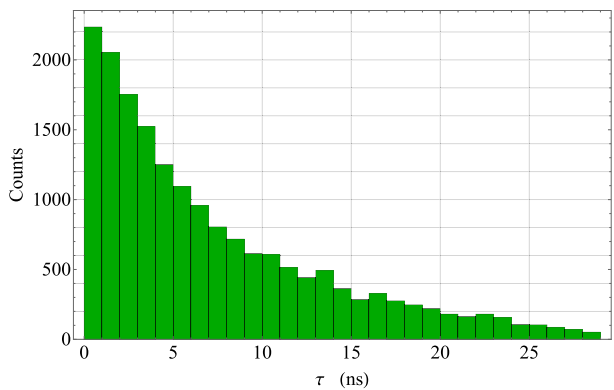

**Fig. 1** Photon counts obtained from an NV fluorescence decay measurement. Histogram generated from the experimental data reported in [13] (cf. Figure 2b)

The experimental configuration utilized in [13] involves a high-pressure high-temperature (HPHT) diamond sample containing a dense concentration of negatively charged NV centers, approximately 18 ppm. The setup includes a custom-made confocal microscope (CM) that focuses the heralded visible photon onto the NV$^-$ centers, resulting in fluorescence photon emission in the 600–800 nm range [13]. The emitted photon is collected and filtered, and its arrival time is recorded by a single-photon avalanche diode (SPAD) linked to a digital oscilloscope. The fluorescence decay data is then used to construct histograms, which form the basis of the analysis in this paper. For the study [13], photon emissions were measured under different pumping power settings (1 mW and 5 mW), with the corresponding heralded photon count rates recorded at 4.5 kcps and 40 kcps, respectively.

The experimental data obtained by the authors of [13] serve as the source for evaluating the performance of statistical estimators for single-photon emission modeling. In this analysis, we focus specifically on the results from the lower pumping power setting, as this data closely follows the characteristics of a standard exponential distribution. The raw data were preliminarily processed to eliminate dark counts, ensuring the photon arrival time distribution asymptotically converged to zero. In Fig. 1, we present a histogram illustrating the distribution of photon arrival times used in the present study. The distribution comprises 17,900 observations.

Following the methodology outlined in Sect. 5, we apply estimators for the exponential distribution to compute the radiative decay time $\tau_R$, which is a key characteristic of single-photon emission. As discussed in Sect. 4, the MVUE is unsuitable for this application because $\tau_R$ corresponds to the scale parameter $\beta$. Therefore, for complete-data estimation, we implement the MLE as defined in Eq. (14), with its corresponding variance given in Eq. (16). Similarly, for Type II censored data, we use the estimator in Eq. (27), along with its variance in Eq. (30). The square root of the variance is employed as the standard error, providing a direct measure of the uncertainty in the parameter estimate, expressed in the same units as the parameter itself.

Additionally, we benchmark our results against those reported in [13], where the radiative decay time for the lower pump setting was determined to be 7.17(14) ns.

**Table 1** Radiative decay time estimates for various levels of censoring. The relative error compares the estimated values to the reference result reported in [13]

| Ratio $r/n$ | Result (ns) | Relative error (%) |
| --- | --- | --- |
| 1 | 7.21(6) | 0.57 |
| 0.99 | 7.28(6) | 1.5 |
| 0.95 | 7.43(6) | 3.6 |
| 0.9 | 7.52(6) | 4.9 |
| 0.85 | 7.56(6) | 5.4 |
| 0.8 | 7.60(6) | 6.0 |
| 0.7 | 7.56(7) | 5.4 |
| 0.6 | 7.46(7) | 4.0 |
| 0.5 | 7.39(8) | 3.1 |
| 0.4 | 7.30(9) | 1.8 |
| 0.3 | 7.3(1) | 1.8 |
| 0.2 | 7.42(12) | 3.5 |
| 0.1 | 7.62(17) | 6.3 |
| 0.05 | 8.02(24) | 12 |
| 0.01 | 9.65(54) | 35 |

This value was obtained by model fitting to the experimental data. In this section, we compare the performance of our statistical estimators with this reference value. To provide a quantitative comparison, we also calculate the relative error, which measures how closely our results align with the reference decay time.

The results of our computations are presented in Table 1 for various ratios of $r/n$, which represent different levels of censoring. Specifically, $r/n = 1$ corresponds to complete-data estimation, while $r/n < 1$ indicates Type II censored data, with the ratio $r/n$ denoting the fraction of the dataset used for estimation.

First, for $r/n = 1$, we obtain the radiative decay time 7.21(6) ns, which lies entirely within the uncertainty range of the reference value 7.17(14) ns, spanning the interval [7.03, 7.31] ns [13]. The estimated value is slightly higher than the benchmark result, but the corresponding standard error is significantly smaller, on the order of $10^{-11}$. The higher estimate of the radiative decay time may be attributed to our simplification of neglecting the non-radiative component of the process. Nevertheless, these results can be regarded as high-quality given the assumptions made in the model.

Next, when censoring is applied by reducing the data set and removing the last 1% of the observations, the estimator's performance declines slightly, though the estimated value still lies within the reference interval. By excluding 1% of the largest ordered statistics, we achieve a result with only 1.5% relative error compared to the benchmark value, indicating the reliable performance of the estimator even with reduced data.

By analyzing the findings presented in Table 1, we can make two key comments. First, as the amount of data used in the estimator decreases, we observe an increase in the standard error, implying that the uncertainty in the estimation process becomes more significant. This trend is monotonic and can be explained by the variance formula in Eq. (30), where the parameter $r$ (the size of the censored sample) appears in the denominator. The second observation concerns the non-monotonic behavior of the estimated value for the radiative decay time. Initially, as $r/n$ decreases, the estimated $\tau_R$ diverges increasingly from the benchmark value, reaching a relative error of 6% at $r/n = 0.8$. However, as $r/n$ is further reduced, the relative error begins to decrease, reaching 1.8% for $r/n = 0.3$ and 0.4. Beyond this point, a further reduction in $r/n$ leads to a rapid decline in estimation quality.

The non-monotonic performance of the estimator can be explained by the characteristics of the experimental data, which were collected under a low pump power setting. As shown in Fig. 1, the distribution is non-monotonic, with noticeable fluctuations in the photon counts, particularly in the tail of the distribution, likely caused by the laser pump. These deviations from the ideal exponential distribution suggest that censoring the data may either improve or worsen the estimation quality. Nonetheless, $r/n = 0.3$ appears to be a threshold value, indicating that at least 30% of the ordered statistics are necessary to reliably estimate the radiative decay time. For $r/n = 0.3$, the estimated value of 7.3 ns still lies within the reference interval [7.03, 7.31] ns, with a standard error of approximately $10^{-10}$ and a relative error of 1.8% compared to the benchmark. These findings demonstrate that even with a limited amount of experimental data, the estimator for the exponential distribution can still provide reliable results.

## 7 Discussion and conclusions

In this paper, by recognizing the disparity in timescales between the non-radiative and radiative processes and modeling the fluorescence emission with a single exponential distribution, we simplify the overall analysis of decay times. This approach allows for the direct use of standard exponential estimators to approximate the radiative decay time $\tau_R$ from experimental data, making it a practical and reliable method for analyzing photon emission events.

Using the statistical properties of the exponential distribution, we were able to directly apply well-known estimators, such as MLE, to the experimental data. In particular, the estimated radiative decay time $\tau_R = 7.21(6)$ ns, obtained with our approach for the complete-data set, lies entirely within the reference interval reported in [13]. Moreover, the corresponding standard error for our result is smaller than that found in the reference study, indicating a lower uncertainty achieved by our method.

Another key finding is the performance of our estimators when applied to Type II censored data. Even with a significant reduction in the data set, the estimators still provided reliable results. For instance, with only 30% of the experimental data, the estimated radiative decay time 7.3 ns remained within the reference interval [7.03, 7.31] ns, demonstrating that our approach can effectively handle even incomplete data. This suggests that the method is well-suited for real-world applications where full data sets may not always be available.

However, it is important to acknowledge the uncertainties and non-monotonic behavior observed in the performance of the estimators as the amount of data is reduced. This behavior can be attributed to the nature of the experimental data, which is not purely exponential. The jumps in photon counts caused by laser pumping introduce deviations from the ideal exponential distribution, impacting the accuracy of the estimates in certain cases. Despite these challenges, our analysis shows that with a sufficient portion of the data, the estimators for the exponential distribution still yield reliable and meaningful results.

While our approach offers practical estimations of radiative decay times, it is important to recognize certain limitations and potential sources of error. The primary limitation arises from the assumption of a single exponential decay model, which may oversimplify the complex nature of experimental data. The real photon emission events often exhibit non-exponential behavior due to factors such as laser pumping, environmental interactions, and experimental noise, leading to deviations from the ideal model and impacting estimation accuracy. Additionally, although our method effectively handles Type II censored data, its reliability diminishes with significantly reduced data sets, as expected for estimators operating with limited input. However, the non-monotonic behavior observed in the performance of estimators, especially due to photon count jumps from laser pumping, further highlights the sensitivity of the method to experimental irregularities. To mitigate these issues, more sophisticated modeling approaches could be adopted, such as incorporating multi-exponential components or advanced statistical models that better capture data deviations. Moreover, employing other estimation techniques, such as Bayesian inference or bootstrapping, may enhance the performance, particularly for incomplete or noisy datasets. Finally, implementing noise reduction strategies and data smoothing algorithms could help improve the quality of decay time estimates, ensuring greater reliability under varying experimental conditions.

Overall, the combination of simplified modeling assumptions and statistical estimators proves to be an effective tool for analyzing single-photon emission events. The ability to handle incomplete data sets while maintaining the quality of estimation positions this method as a practical approach for future studies in quantum optics and photon emission modeling.

**Acknowledgements** The experimental data used in this study were obtained with the consent of the authors from the experiment reported in Ref. [13].

**Data availability** The data supporting the results reported in this paper can be obtained from the corresponding author, provided that consent is granted by the authors of Ref. [13].

## Declarations

**Conflict of interest** The authors declare that they have no conflict of interest.

# References

1. Piro, N., Rohde, F., Schuck, C., Almendros, M., Huwer, J., Ghosh, J., Haase, A., Hennrich, M., Dubin, F., Eschner, J.: Heralded single-photon absorption by a single atom. Nat. Phys. **7**, 17–20 (2011)
2. Aharonovich, I., Englund, D., Toth, M.: Solid-state single-photon emitters. Nat. Photon. **10**, 631–641 (2016)
3. Brito, J., Kucera, S., Eich, P., Mueller, P., Eschner, J.: Doubly heralded single-photon absorption by a single atom. Appl. Phys. B **122**, 36 (2016)
4. Wang, Y., Um, M., Zhang, J., An, S., Lyu, M., Zhang, J.-N., Duan, L.-M., Yum, D., Kim, K.: Single-qubit quantum memory exceeding ten-minute coherence time. Nat. Photon. **11**, 646–650 (2017)
5. Pirandola, S., Andersen, U.L., Banchi, L., Berta, M., Bunandar, D., Colbeck, R., Englund, D., Gehring, T., Lupo, C., Ottaviani, C., Pereira, J., Razavi, M., Shaari, J.S., Tomamichel, M., Usenko, V.C., Vallone, G., Villoresi, P., Wallden, P.: Advances in quantum cryptography. Adv. Opt. Photon. **12**, 1012–1236 (2020)
6. Meda, A., Losero, E., Samantaray, N., Scarimuto, F., Pradyumna, S., Avella, A., Ruo-Berchera, I., Genovese, M.: Photon-number correlation for quantum enhanced imaging and sensing. J. Opt. **19**, 094002 (2017)
7. Camphausen, R., Perna, A.S., Cuevas, Á., Demuth, A., Chillón, J.A., Gräfe, M., Steinlechner, F., Pruneri, V.: Fast quantum-enhanced imaging with visible-wavelength entangled photons. Opt. Express **31**, 6039–6050 (2023)
8. Scully, M.O., Zubairy, M.S.: Quantum Optics. Cambridge University Press, Cambridge (1997)
9. Meystre, P., Scully, M.O.: Quantum Optics. Taming the Quantum, Springer, Berlin, Germany (2021)
10. Choi, J.R.: The decay properties of a single-photon in linear media. Chin. J. Phys. **41**, 257–266 (2003)
11. Bennett, A.J., Unitt, D.C., See, P., Shields, A.J., Atkinson, P., Cooper, K., Ritchie, D.A.: Electrical control of the uncertainty in the time of single photon emission events. Phys. Rev. B **72**, 033316 (2005)
12. Li, Q., Orcutt, K., Cook, R.L., Sabines-Chesterking, J., Tong, A.L., Schlau-Cohen, G.S., Zhang, X., Fleming, G.R., Whaley, K.B.: Single-photon absorption and emission from a natural photosynthetic complex. Nature **619**, 300–304 (2023)
13. Gieysztor, M., Misiaszek, M., van der Veen, J., Gawlik, W., Jelezko, F., Kolenderski, P.: Interaction of a heralded single photon with nitrogen-vacancy centers in a diamond. Opt. Express **29**, 564–570 (2021)
14. Gieysztor, M., Misiaszek, M., van der Veen, J., Gawlik, W., Jelezko, F., Kolenderski, P.: Supplementary document for Interaction of a heralded single photon with nitrogen-vacancy centers in diamond. (2021). https://doi.org/10.6084/m9.figshare.13383020.v3
15. Mood, A.M., Graybill, F.A., Boes, D.C.: Introduction to the Theory of Statistics. McGraw-Hill, New York (1974)
16. Hogg, R.V., Craig, A.T.: Introduction to Mathematical Statistics, 4th edn. Macmillan, New York, NY, USA (1978)
17. Ross, S.M.: Introduction to Probability Models, 9th edn. Elsevier, London, UK (2007)
18. Nelson, W., Hahn, G.J.: Linear estimation of a regression relationship from censored data-part II best linear unbiased estimation and theory. Technometrics **15**, 133–150 (1973)
19. Czerwinski, A., Czerwinska, K.: Statistical analysis of the photon loss in fiber-optic communication. Photonics **9**, 568 (2022)
20. Fasel, S., Alibart, O., Tanzilli, S., Baldi, P., Beveratos, A., Gisin, N.: Zbinden, H,: High-quality asynchronous heralded single-photon source at telecom wavelength. New J. Phys. **6**, 163 (2004)
21. Collins, M.J., Xiong, C., Rey, I.H., Vo, T.D., He, J., Shahnia, S., Reardon, C., Krauss, T.F., Steel, M.J., Clark, A.S., Eggleton, B.J.: Integrated spatial multiplexing of heralded single-photon sources. Nat. Commun. **4**, 2582 (2013)
22. Specht, H.P., Noelleke, C., Reiserer, A., Uphoff, M., Figueroa, E., Ritter, S., Rempe, G.: A single-atom quantum memory. Nature **473**, 190–193 (2011)
23. Neumann, P., Jakobi, I., Dolde, F., Burk, C., Reuter, R., Waldherr, G., Honert, J., Wolf, T., Brunner, A., Shim, J.H., Suter, D., Sumiya, H., Isoya, J., Wrachtrup, J.: High-precision nanoscale temperature sensing using single defects in diamond. Nano Lett. **13**, 6 (2013)
24. Doherty, M.W., Manson, N.B., Delaney, P., Jelezko, F., Wrachtrup, J., Hollenberg, L.C.: The nitrogen-vacancy colour centre in diamond. Phys. Rep. **528**, 1–45 (2013)

25. Schabikowski, M., Wojciechowski, A., Mitura-Nowak, M., Mrózek, M., Kruk, A., Rajchel, B., Gawlik, W., Marszałek, M.: Optical characterization of nitrogen-vacancy centers created by proton implantation in diamond. Acta Phys. Pol., A **137**, 9–12 (2020)
26. Mrózek, M., Schabikowski, M., Mitura-Nowak, M., Lekki, J., Marszałek, M., Wojciechowski, A.M., Gawlik, W.: Nitrogen-vacancy color centers created by proton implantation in a diamond. Materials **14**, 833 (2021)
27. Batalov, A., Zierl, C., Gaebel, T., Neumann, P., Chan, I.-Y., Balasubramanian, G., Hemmer, P.R., Jelezko, F., Wrachtrup, J.: Temporal coherence of photons emitted by single nitrogen-vacancy defect centers in diamond using optical rabi-oscillations. Phys. Rev. Lett. **100**, 077401 (2008)
28. Romeu, J.L.: Censored Data. Selected Topics in Assurance Related Technologies **11**, 1–8 (2004)
29. Witte, R.S., Witte, J.S.: Statistics. Wiley (2017)
30. Bohm, G., Zech, G.: Introduction to Statistics and Data Analysis for Physicists, vol. 1. Hamburg, Germany, Desy Book (2010)
31. Czerwinski, A., Szlachetka, J.: Efficiency of photonic state tomography affected by fiber attenuation. Phys. Rev. A **105**, 062437 (2022)
32. Heumann, Ch., Shalabh, M.S.: Introduction to Statistics and Data Analysis. Cham, Switzerland, Springer Nature (2016)
33. David, H.A., Nagaraja, H.N.: Order Statistics. Wiley, Hoboken (2003)
34. Lehmann, E.L., Scheffé, H.: Completeness, similar regions, and unbiased estimation. I. Sankhyā **10**, 305–340 (1950)
35. Lehmann, E.L., Scheffé, H.: Completeness, similar regions, and unbiased estimation. II. Sankhyā **15**, 219–236 (1955)
36. Cramér, H.: Mathematical Methods of Statistics. Princeton University Press, Princeton, NJ, USA (1946)
37. Cohen, A.C.: Simplified estimators for the normal distribution when samples are singlycensored or truncated. Technometrics **1**, 217–237 (1959)
38. Balakrishnan, N., Cohen, A.C.: Order Statistics and Inference. Estimation Methods, 1st ed. Academic Press Inc.: London, UK (1991)
39. Smith, P.J.: Analysis of Failure and Survival Data, 1st edn. Chapman and Hall, New York, NY, USA (2002)
40. Mahmoud, M.R., Sultan, K.S., Saleh, K.M.: Progressively censored data from the linear exponential distribution: moments and estimation. Int. J. Stat. **64**, 199–215 (2006)
41. Henderson, C.R.: Best linear unbiased estimation and prediction under a selection model. Biometrics **31**, 423–447 (1975)
42. Andronov, A., Spiridovska, N., Santalova, D.: On a Parametric Estimation for a Convolution of Exponential Densities. In: Pilz, J., Melas, V.B., Bathke, A. (eds.) Statistical Modeling and Simulation for Experimental Design and Machine Learning Applications, pp. 181–195. SimStat (2023)
43. Collins, J.: Non-Radiative Processes in Crystals and in Nanocrystals. ECS J. Solid State Sci. Technol. **5**, 3170 (2016)

## Authors and Affiliations

**Artur Czerwinski[1]**  **· Katarzyna Czerwinska[2]**  **· Xiangji Cai[3]**  **· Asad Ali[4]**  **· Hashir Kuniyil[4]**  **· Atta ur Rahman[5]**  **· Saif Al-Kuwari[4]**  **· Saeed Haddadi[6]**

✉ Artur Czerwinski
aczerwin@umk.pl

Xiangji Cai
xiangjicai@foxmail.com

Asad Ali
asal68826@hbku.edu.qa

Hashir Kuniyil
hkuniyil@hbku.edu.qa

Atta ur Rahman
attaphy57@mails.ucas.ac.cn

Saif Al-Kuwari
smalkuwari@hbku.edu.qa

Saeed Haddadi
haddadi@ipm.ir

1　Institute of Physics, Faculty of Physics, Astronomy and Informatics, Nicolaus Copernicus University in Torun, ul. Grudziadzka 5, 87-100 Torun, Poland

2　Foundation for Responsible Education and Upbringing "Ad Veritatem", ul. Piskorskiej 4/4, 87-100 Torun, Poland

3　School of Science, Shandong Jianzhu University, Jinan 250101, China

4　Qatar Center for Quantum Computing, College of Science and Engineering, Hamad Bin Khalifa University, Doha, Qatar

5　School of Physical Sciences, University of Chinese Academy of Sciences, Yuquan Road 19A, Beijing 100049, People's Republic of China

6　School of Particles and Accelerators, Institute for Research in Fundamental Sciences (IPM), P.O. Box 19395–5531, Tehran, Iran